\documentstyle[12pt]{article}
\begin{document}
\topmargin=-.5in
\oddsidemargin=.1in
\evensidemargin=.1in
\vsize=23.5cm
\hsize=16cm
\textheight=23.0cm
\textwidth=16cm
\baselineskip=24pt
\thispagestyle{empty}

\hfill{ITP-SB-98-41}\\

\smallskip
\hfill{ June 1998}

\vspace{0.5in}

\centerline{\Large \bf Peculiar Macroscopic And Microscopic} 

\centerline{\Large \bf Properties Of A Fractional Quantum Hall Layer}

\vspace{.85in}

{{\baselineskip=16pt
\centerline{\large Alfred Scharff Goldhaber}
\bigskip
\centerline{\it Institute for Theoretical Physics}
\centerline{\it State University of New York}
\centerline{\it Stony Brook, NY 11794-3840}}

\vspace{.5in}

\centerline{\Large Abstract}

\vspace{.5in}

 At a surface between electromagnetic media
the Maxwell equations allow either the usual boundary conditions,
or exactly one alternative: continuity of  
${\bf E_\perp}$,
 ${\bf H_\perp}$, ${\bf D_\parallel}$, ${\bf B_\parallel}$.
These `flipped' conditions on the top and bottom surfaces of an FQH layer capture
all its known static electromagnetic properties  
and so may be considered a deduction from microscopic quantum theory, yet are
unobtainable in any realistic, purely classical model.  An unrealistic model
with free magnetic monopole currents illustrates this.

 At the microscopic
level, the FQH system is a laboratory for
the particle concept.  Identifying quasiparticles in terms of kinematics
or in terms of asymptotic states gives two different perspectives.  The
kinematics exhibits exclusion rules similar to those in exactly solvable
models for quantum systems in one space dimension.  An effort here to tie some
of these aspects together may be useful as a foundation for a future more
comprehensive assessment of the roles and limitations of quasiparticles.

\newpage

Both
the original \cite {K} and  the 
fractional quantum Hall  effects \cite{TS} are striking illustrations
of ways in which quantum phenomena can violate classical intuition.
In particular, the fractional effect  [FQHE] is remarkable for many reasons,
not least that in a very short time it found its `standard model' in the
composite-fermion picture \cite{J}, which unifies Laughlin's original description
of simple Hall fractions $\nu = \frac{1}{2p+1}$ \cite{Laugh} with a host of other
observed phenomena in the fractional Hall domain, as well as earlier
understanding \cite {Laugh0} of the integer quantum Hall effect. 
Perhaps it is because progress in microscopic theory has been so rapid that a
familiar stage from previous studies of macroscopic systems -- phenomenological
description in terms of an electromagnetic medium -- appears to have been
skipped.  Another possible reason is that the thickness of the layer (${\cal 
O} (500 \AA )$) is small enough that attempting a macroscopic description
might have been viewed as a risky enterprise.  

One purpose of the present work is to show that indeed there is a unique, 
internally consistent variation on usual macroscopic descriptions of
electromagnetic media which does fit the FQHE perfectly.  This description
cannot be reproduced by any realistic classical model, although a
model involving copious supplies of freely moving magnetic monopoles (objects
yet to be observed) can account for most of the desired features, thus 
underlining
the exotic character of the FQHE.  Another feature  with roots in classical
physics which remains essential in quantum considerations is the concept of a
particle.  Quasiparticle excitations are used to describe phenomena for FQH just
as for many other systems, but there are a number of subtleties about how and to
what extent a description like one associated with familiar elementary particles
remains valid in this setting. The second goal of this work is tying together
some relevant observations scattered in the literature, as a way of laying a
foundation for a comprehensive discussion of quasiparticles, their nature and
their fragility.

Jain noted \cite{J,GJ} that the FQHE can be described as combining a familiar
property, renormalization of local quasiparticle charge by polarization of the
medium, with an entirely novel property extending even beyond the fractional Hall
plateaux, renormalization of the perpendicular magnetic field and tangential
electric field inside the medium with respect to values in the external regions
immediately adjacent to the surfaces of the Hall layer.  Such multiplicative
renormalization of external electromagnetic field components in turn was a
development of the notion that in the medium there is gauge coupling not only
to the electromagnetic field but also to a Chern-Simons field whose strength
is proportional to the electron density \cite {CS1,CS2}, yielding an additive
renormalization of the effective field.
 These results for FQHE
may be summarized by the statements:

\noindent
1)  The Gauss-law charge of a quasiparticle, measured by its electric field far
away, is $e^\star$, where the ratio $e/e^\star = 2pn\pm 1$ is related to the
Hall fraction
$\nu = \frac{n}{2pn\pm 1}$.

\noindent
2)  Quasiparticles move in the presence of an  electromagnetic field 
as
if they carried electric charge
$e^\star$ -- or equally well as if they carried a charge $e$, but in
 fields
${\bf B}^\star _\perp=e^\star{\bf B}_\perp /e$ and  ${\bf E}^\star _\parallel
=e^\star {\bf E}_\parallel/e$.

Further, it is well accepted that 

\noindent
3)  If an electron were gently inserted into the surface of a Hall layer, it
would break up into $2pn\pm 1$ quasiparticles \cite{Su}.

A naive effort to describe 
the Hall layer as a conventional dielectric medium fails at once.  To take the
most elementary aspect, far from a charge located in a thin dielectric layer there
is no renormalization of that charge by the dielectric constant of the medium,
because the surface charge which together with the local charge makes a total
$e$ is itself localized quite close to the particle.
By the same token, property 2) also does not hold.  Is there any other option? 
The usual boundary
 conditions for the electromagnetic fields at a surface between two different 
media
are continuity of
${\bf D}_\perp , {\bf B}_\perp, {\bf E}_\parallel,{\bf H}_\parallel$.  These
conditions follow directly from a classical model of the medium as
a collection of small electric and magnetic dipoles whose orientation
and/or dipole moment magnitude is affected by the applied electric
or magnetic field.  

However,
it is consistent with the Maxwell equations to interchange simultaneously the
roles of ${\bf E}$ and ${\bf D}$, ${\bf B}$ and ${\bf H}$, imposing continuity of
${\bf E}_\perp , {\bf H}_\perp, {\bf D}_\parallel, {\bf B}_\parallel$.  Note
that just interchange of one pair would not be consistent.  An easy way to see
this is that such mixed boundary conditions would violate the duality-rotation
symmetry of the Maxwell equations (discussed more below) under which ${\bf E
\rightarrow H}$  and ${\bf D \rightarrow B}$.  
 Thus, there is a
 unique alternative to
consider as a possible description of the FQHE.  Let us test it.

Assume
the dielectric constant $\epsilon$ and magnetic permeability $\mu$ are given by
$\epsilon = \mu ^{-1} = e/e^\star$.  Then continuity of $E_\perp$ at the top
and bottom surfaces assures that the total electric flux coming out of a charge
placed in the layer corresponds to a charge $e^\star$, not $e$, thus confirming
1).  Continuity of ${\bf H}_\perp$ yields a perpendicular magnetic field inside
the layer of magnitude $\mu B$, , and continuity of ${\bf D}_\parallel$ implies
a parallel electric field inside of magnitude $E/\epsilon$, confirming 2).
If an electron descends into the Hall layer, total local charge $e$ must be
conserved, but inside the layer each individual charge is $e^\star$.  In a
conventional insulator, the electron would leave part of its charge on the
surface, but in
our
context, where distributing the residual charge in the surface of the
layer is not possible,   
the only way to achieve local implementation of the conservation law is by
generation of
$2pn\pm 1$ quasiparticles.  Thus, even fact 3) follows from this novel set of
constitutive relations.  Note that, although the treatment here is not
explicitly quantum-mechanical, local conservation of electric charge requires
quantization of $\epsilon$ at integer values, as otherwise conservation of
charge through breakup would not be possible \cite{Su}.  The fact that only
odd integers are allowed follows because at the core of each quasiparticle 
must be a fermion, and by a superselection rule an odd number of fermions may
not turn into an even number \cite{WWW}.

In other condensed-matter settings, quasiparticles often
are described as dressed
electrons, meaning they have renormalized local charge, but unrenormalized
Aharonov-Bohm  or Lorentz-force charge $e$, i.e., 
in a specified electromagnetic field a quasiparticle experiences exactly the
same influence as would a free electron in the same field configuration. 
Clearly that assumption is intrinsic in the above deductions also.  However, this
leads to a potentially disturbing result, that penetration of an electron into the
layer, with resulting breakup, does not conserve Aharonov-Bohm charge.  For
a gauge shift $\lambda$ nominally uniform throughout space, this result can be
made consistent with usual considerations of gauge invariance if we insist 
that inside the layer the corresponding shift is
$\lambda^\star=\lambda/\epsilon$.  

Now this definition is nothing more than
a convenient bookkeeping choice, but if we go on to consider the spatial
variation of $\lambda$ in a direction parallel to the
Hall layer, then we are forced by the boundary conditions to insist that this
variation, which is related to the tangential part of the vector
potential ${\bf A}$, must be modified by the factor $\epsilon^{-1}$ inside the
layer compared to outside. The reason is that in $A_0 = 0$ gauge the time
derivative of ${\bf A}$ gives the electric field ${\bf E}$, and therefore the
tangential component of ${\bf A}$ must be renormalized exactly as the
tangential component of ${\bf E}$.  Thus, the only consistent choice in
this scheme is one in which an otherwise continuous $\lambda$ defined everywhere
is renormalized to $\lambda^\star$ inside the layer.  The same conclusion
follows if we pay attention to the discontinuity in ${\bf B_\perp}$, emphasizing
the necessity already discussed of interchanging both magnetic and electric field
boundary conditions together to assure mathematical consistency of the
description.

The consequence of $\lambda$ renormalization is that $\sum \lambda_i e_i$ for
all charges must be time independent when an electron passes into the layer,
making non-conservation of AB charge not only consistent but actually
required in this context.

It is
interesting to ask if there might be a strictly classical model which would
exhibit all or some of these features.  As electric charge should be continuously
adjustable in such a model, we would not expect the quantization of $e/e^\star$
which characterizes FQHE, and therefore would not
expect breakup as an electron enters the Hall medium.  As we have seen already,
that implies a failure of local conservation of electric charge.  Recognizing
these limitations, let us see whether a model reproducing features 1) and 2) is
possible.
  For
a classical medium, there is no escape from the usual boundary conditions, but one
may describe electric charge in an exotic way:  Consider a solenoid carrying not
ordinary electric current but rather magnetic-monopole current.  Then one end of
the solenoid looks very much like an electric charge.  Suppose the 
solenoid is very long (so that interactions involving the other end of the
solenoid may be ignored) and lies in a direction parallel to the Hall
layer.  We may call the nearby end an `artificial' charge.

Electric-magnetic duality tells us that monopoles respond to ${\bf H}$ and ${\bf
D}$ in the way that electric charges respond to ${\bf E}$ and ${\bf B}$.  An
early example of this conclusion came in the elegant energy-conservation argument
of Kittel and Manoliu that a monopole interacting with a ferromagnet cannot
respond to ${\bf B}$ \cite{KM}.  Another way to see this is to require that
the Maxwell equations in a medium should exhibit duality symmetry.  In that
case, the usual boundary conditions evidently are consistent with duality only if
the field correspondences are as indicated.  To complete the correspondence, we
must identify a free monopole as an unrenormalized source of ${\bf B}$ and ${\bf
E}$ in just the way that a free electric charge is an unrenormalized source of
${\bf D}$ and
${\bf H}$.    These
interchanges mean that a medium which is dielectric for ordinary charges 
and diamagnetic for artifical magnetic monopoles (i.e., ends of ordinary
solenoids) is `para-electric' (antiscreening) for artificial charges
and paramagnetic for true magnetic monopoles.  For this
reason, classical considerations of the constitutive relations already
imply that
renormalizations of true magnetic and electric charges must be reciprocal to each
other,  making renormalization consistent with Dirac's quantum condition
\cite{Dirac} on the product  of electric and magnetic charge.  This issue was
a source of confusion at an earlier stage in quantum-theoretical studies of
monopoles
\cite {JS}, but in that context was clarified long ago \cite {SC}.

The above considerations imply that we want to allow our
artificial charges to move in a layer with usual dielectric constant $\epsilon '
=\mu '^{-1}= e^\star /e < 1$, i.e., a para-electric and
paramagnetic medium for ordinary charges.  This means that when the solenoid
is submerged in the Hall layer it will have a $D$ flux through its end smaller
by a factor $e^\star /e$ than when the solenoid is outside the layer, and hence
the effective charge of an isolated quasiparticle will be $e^\star$, just as
required.  It is easy to check that the other requirements also are obeyed.
Of course, a para-electric medium is itself peculiar, but perhaps one may be
forgiven for assuming such a system in a hypothetical world where solenoids of
magnetic current are available!  

	As mentioned, the above description implies that when the solenoid
descends into the layer its electric flux is bleached, with flux lines
disappearing all along the length of the solenoid simultaneously.  That makes
the descent of a charge into the layer, and its resulting attenuation, a highly
nonlocal process.   Of course, this classical model is extraordinarily
different from the actual quantum system, but the nontriviality of the
conversion from `electron' to `quasiparticle' is found again.  For entrance of an
actual electron into the Hall layer, electric charge is conserved but particle
number is not, while the classical model conserves particle (i.e.,
solenoid end) number, but lets charge leak to the other end of the solenoid. 
The model is less flexible than the constitutive relations plus boundary
conditions, because it would give the same peculiar properties at a second 
boundary between two non-FQHE layers, contrary to experience.  Nevertheless,
the model may be helpful for visualizing the
notion of flipped boundary conditions, and understanding why they could never
occur for conventional classical systems.

So far we have seen how the constitutive relations and surface boundary
conditions reproduce known results, but one might wonder if they can give any new
information.  One obvious question to address is whether the Hall layer exhibits
new trapped modes, which in turn might be detected by scattering experiments. 
For electromagnetic waves, as mentioned just above, the Hall layer is equivalent
by duality to a para-electric, paramagnetic layer.  Such a system has no trapped
modes, because critical internal reflection requires refractive index greater
than unity inside, and in this case we have (relative) refractive index exactly
equal to unity.  Furthermore, the fact that the layer must be quite thin on the
wavelength scale relevant to the FQHE means that even reflection of external
waves must be strongly suppressed.  Besides possible optical modes, one might
wonder if there could be a longitudinal mode, but any such mode must have finite
mass because of the incompressibility of the FQH ground state in a specified
perpendicular magnetic field.  Of course, for compressible states, as exist near
$\nu=\frac{1}{2}$, there can be longitudinal modes, and consideration of these
has led to the suggestion that there may be  modifications of the
simple composite-fermion Fermi surface expected if these modes
were ignored \cite {HLR}.  Recent studies have focused on these longitudinal
`plasmon' modes to obtain remarkably detailed analytic results for both the
compressible regime and the incompressible regime (where the plasmon has an
effective mass) \cite {SM}. 

Of course, the defining property of the FQH layer is the Hall effect.  How is
that incorporated into this formulation?   If we label the direction normal to
the layer as $\hat z$, then in view of the boundary conditions the Hall current
should be given by $${\bf J} = \nu \frac {e^2}{h}f(z)\hat z \times {\bf D}.$$
Here the integral of $f(z)$ over the thickness $t$ of the layer is normalized to
unity, so that $f$, which reflects the (rigid) distribution of the contributing
electron wave function in the $z$ direction, 
 scales inversely with $t$.  For one example, associated with a plane
electromagnetic wave propagating parallel to the layer (with its $B$ perpendicular
to the layer), there will be an induced oscillating longitudinal electric current,
and by local charge conservation an associated oscillating electric charge
density.  These currents should be at most weak sources of electromagnetic
radiation, because they are longitudinal except at edges and surfaces of the
sample.  On the other hand, an incoming wave with normal incidence would induce
strong dipole radiation from the sample as a whole, having polarization
rotated by
$\pi/2$ with respect to that of the incident wave.  

The dielectric response function of any medium should depend on
frequency and wavenumber.  Provided $\epsilon$ and $\mu$ approach unity
 as the frequency rises above some critical
value, the unique description proposed here should remain in
agreement with observation.  In particular, it should be possible to describe
the full frequency and wavenumber dependence of $\epsilon$, including a peak
in Raman response found for the $\nu=1/3$ state \cite {Pin}.  

Another aspect which deserves attention is the boundary conditions on the
edge surfaces of the FQH layer, which of course are quite narrow compared to
the top and bottom surfaces.  At an edge one must have (partially) the traditional
conditions.  This is immediately clear for ${\bf D_{\perp}}$ and ${\bf
B_{\perp}}$ and for the components of ${\bf H_{\parallel}}$ and ${\bf
E_{\parallel}}$ lying perpendicular to the Hall plane.  On the other hand,
consistency with conditions on the top and bottom surfaces also imply that the
components of
${\bf D_{\parallel}}$ and ${\bf B_{\parallel}}$ lying in the Hall plane 
should be continuous.  At first sight these hybrid conditions might seem
unsatisfactory.  However, because the strong external magnetic field 
perpendicular to the layer determines
a special direction, these requirements are no more peculiar than the
familiar appearance of a complicated dielectric tensor in some anisotropic
medium.  

If anything, it is remarkable that in this case the electric and
magnetic  susceptibilities are just
scalars, with the anisotropy entirely described by the boundary conditions. 
One might ask whether the hybrid boundary conditions might lead
to some interesting behavior near the edge of the sample.  It seems plausible
that the answer might be affirmative.  Continuity of ${\bf D_{\perp}}$ suggests
that a single electron entering might proceed as a quasiparticle into the medium,
leaving the remainder of the charge to populate edge states, and thus mimicking
-- on the edge only -- the induced charge found on the surfaces of a traditional
insulator.  Whether this possibility gives any useful insight about edge states
remains to be seen.
   
Even though the emphasis above has been on a macroscopic description, we were
compelled to focus on quasiparticles, as the objects which act as sources of the
fields described, fields which in turn influence the motion of the particles. 
Therefore it may be worth noting that the term `quasiparticle' is used in two
related but distinct senses.  The one of primary interest here is as an
excitation which is stable in isolation, having well-defined properties
determining its (weak) long-range interactions with other quasiparticles.  The
related meaning is as an entity appearing in an effective dynamics divided into a
`kinetic' part for which these quasiparticles move without interaction, and an
`interaction' part which is weak enough to be treated by perturbative
techniques.  

Composite fermions are quasiparticles in this second sense, and
therefore are spin-half particles with intrinsic electric charge equal to that of
an electron, and ordinary Fermi-Dirac statistics.  However, the weak
residual-interaction part of the dynamics implies that the phenomenological
quasiparticles associated with the first meaning have fractional local charge
\cite{Laugh}, exhibit a dynamically induced long-range interaction \cite{BH} 
associated with the concept of `fractional statistics' \cite {LM,W}, and possess
a well-defined fractional localized contribution to the spin \cite{GJ,Sondhi}. 
Failure to distinguish between these two meanings of a single term can easily
lead to confusion.  This discussion brings up a subtle discontinuity between the
macroscopic and microscopic descriptions.  Composite fermions, being just
strongly-correlated electrons, have conserved Aharonov-Bohm charge $e$ just like
electrons.  Thus in the composite-fermion picture the addition of an electron
not only produces an increase of AB charge for the several quasiparticles
produced, but also a decrease in the number of composite fermions, and hence the
AB charge, of the ground state on which the quasiparticle excitations are built
\cite{GJ}.  Because this decrease is spread uniformly over the entire Hall
layer, it represents an intrinsically nonlocal aspect of the electron
`splitting', and by the same token becomes unobservable even in principle once
the thermodynamic limit holds.   

The subtle limitations on the applicability of the quasiparticle concept
illustrated by this discussion are part of a pattern which is getting
increasing recognition.  The description of quasiparticle dynamics tends to
involve
 constraints  on allowed configurations
which are more complex than the simple Bose-Einstein
or Fermi-Dirac statistics \cite{HAL}.  Especially notable, and relevant to
our case, are generalized exclusion rules
found in exactly soluble quantum problems in one space dimension, where adding a
quasiparticle  or quasihole to 
a configuration modifies the number of available hole and particle states, and
not just the ones which have been emptied or filled
\cite{MCC}.  A pertinent example is the resonance observed in Raman scattering
for the $\nu = \frac{1}{3}$ system \cite{Pin}, which one would like to describe
as a quasiparticle-quasihole pair.  The most obvious excitation at small spatial
momentum transfer is one in which the created quasiparticle sits geometrically
on top of the hole formed by the excitation.  However, because all the states
under consideration lie in the lowest Landau level in the external magnetic
field, Kohn's theorem \cite{Kohn} forbids this pure dipole configuration to
occur, and thus gives directly a constraint on which particle-hole
configurations can form.  It seems possible that more detailed examination of
the allowed excitations will show a strong correspondence with what has been
found in exactly solvable models, but perhaps also some significant
differences, as it is by no means obvious that the FQH dynamics constitute an
exactly solvable system.

We may conclude that the long-wavelength physics of the
FQHE is captured by introducing the
familiar constitutive relations for the dielectric and diamagnetic response of
the Hall layer, \underline{except} that the conventional continuity conditions 
on the electromagnetic fields at the top and
bottom surfaces are interchanged between perpendicular and
parallel components  of the fields.  No simple classical system
duplicates this (although an unrealistic
 model with artificial electric
charges comes close).  Thus, the flipped surface conditions
express in macroscopic terms the quantum `magic' of the fractional quantum Hall
effect.  At the same time, phenomena associated with excitations,
including addition of an electron to the system as well as bosonic excitations
which evoke description in terms of particle-hole pairs, give a fascinating
glimpse for a real system of ways in which the particle concept begins to
fray when it is pushed very hard.  Work is underway on exploring this
phenomenon, and its connections with behaviors for exactly soluble models in one
space dimension \cite {GJMP}. 

This work was supported in part by the National Science Foundation.  I thank
Jainendra Jain and Barry McCoy
for discussions, and Steven Kivelson for emphasizing some time ago
that a conventional electromagnetic medium fails to describe the fractional
quantum Hall effect.

\end{document}